\documentstyle[11pt,aaspp4]{article}

\begin{document}

\title{Directed Relativistic Blast Wave} 

\author{Andrei Gruzinov}

\affil{CCPP, Physics, New York University, 4 Washington Place, New York, NY 10003}

\begin{abstract}

A spherically symmetrical ultra-relativistic blast wave is not an attractor of a generic asymmetric explosion. Spherical symmetry is reached only by the time the blast wave slows down to non-relativistic velocities, when the Sedov-Taylor-von Neumann attractor solution sets in. We show however, that a directed relativistic explosion, with the explosion momentum close to the explosion energy, produces a blast wave with a universal intermediate asymptotic -- a selfsimilar directed ultra-relativistic blast wave. This universality might be of interest for the astrophysics of gamma-ray burst afterglows.

\end{abstract}

\section{Relativistic Blast Waves and Gamma-Ray Burst Afterglows}

We give a selfsimilar solution describing the opening of a narrow ultra-relativistic blast wave. This solution is an attractor (intermediate asymptotic) of a generic directed explosion. Here the ``directed explosion'' means an explosion with momentum $P$ nearly equal to the explosion energy $E$ ($c=1$ here and below). Qualitatively, the directed blast wave solution has been already discussed by Rhoads (1999).

A non-directed explosion (with $E-P\lesssim E$) does not have a universal intermediate asymptotic during the ultra-relativistic stage. The Blandford-McKee (1976) selfsimilar solution is an intermediate asymptotic only for a spherically symmetric explosion (Gruzinov 2000). A generic non-directed explosion acquires spherical symmetry only at the non-relativistic stage, when it asymptotes to the Sedov-Taylor-von Neumann attractor solution.

Qualitatively, the non-universality of the Blandford-McKee selfsimilar solution follows from causality. The shock of the Blandford-McKee solution moves with Lorentz factor $\Gamma \propto t^{-3/2}$. Consider a light signal propagating along the shock, starting at a polar angle $\theta =0$. Then $(dr/dt)^2+(rd\theta /dt)^2=1$, which, for $\Gamma \gg 1$, gives $\theta ={2\over 3}\Gamma ^{-1}$. Two regions of the shock, separated by angle $\theta$, do not talk to each other until the blast wave slows down to $\Gamma \sim \theta ^{-1}$. 

On the other hand, the selfsimilar solution derived here has a shock of opening angle $\theta \sim  \Gamma^{-1}$. The angular structure of the blast wave is not imposed by hand -- it appears dynamically, from an arbitrary initial state (satisfying the directed explosion requirement $E-P\ll E$). Numerical simulations confirm the universality of the resulting selfsimilar solution.

It is thought that ultra-relativistic blast waves are responsible for the observed gamma-ray burst afterglows (Piran 1999). Universality of the blast waves resulting from directed explosions should be of interest for astrophysics (Rhoads 1999).

We give the main result in \S2, derive it in \S3, and describe the numerical simulations in \S4.

\begin{figure}
\plottwo{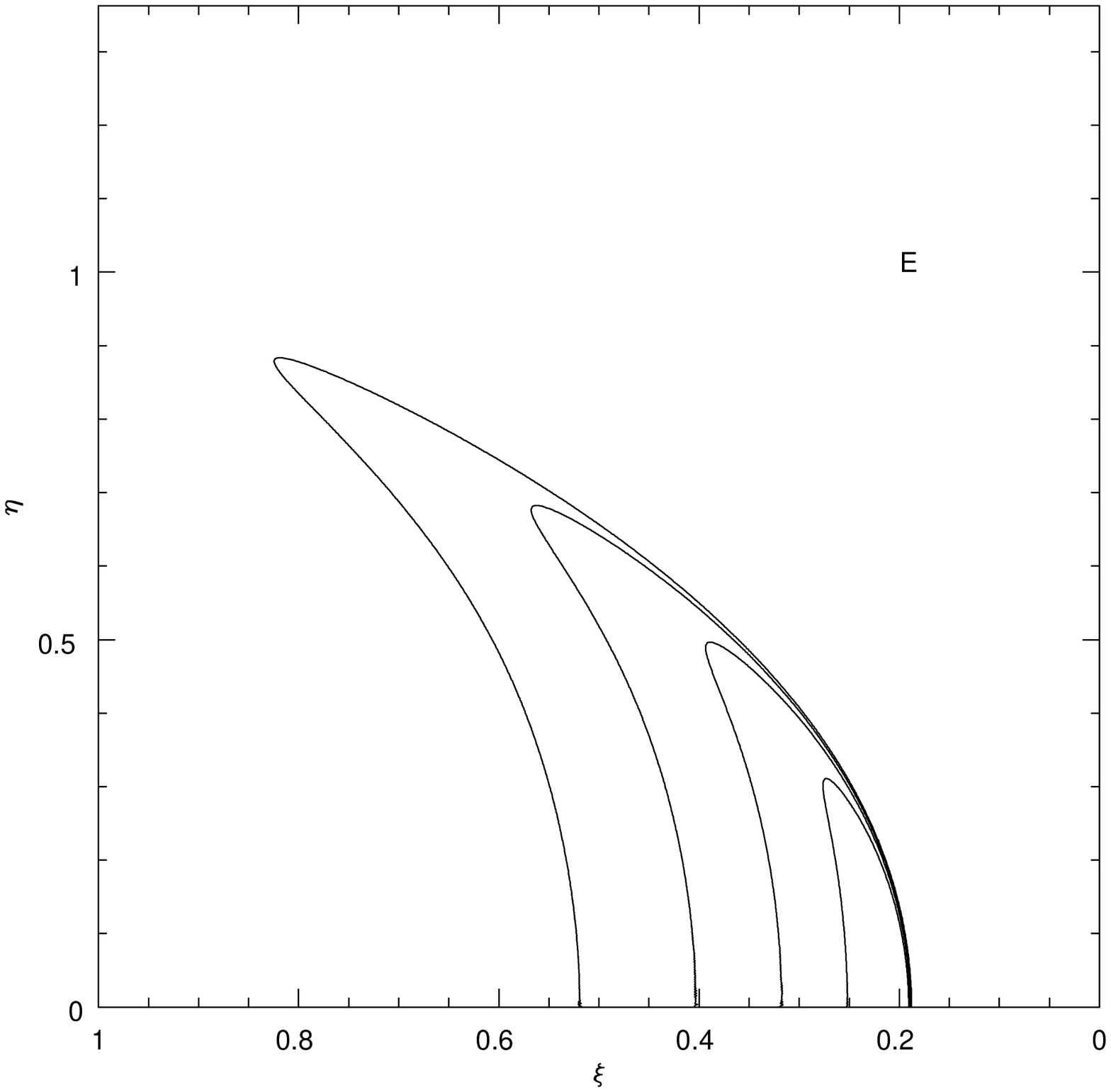}{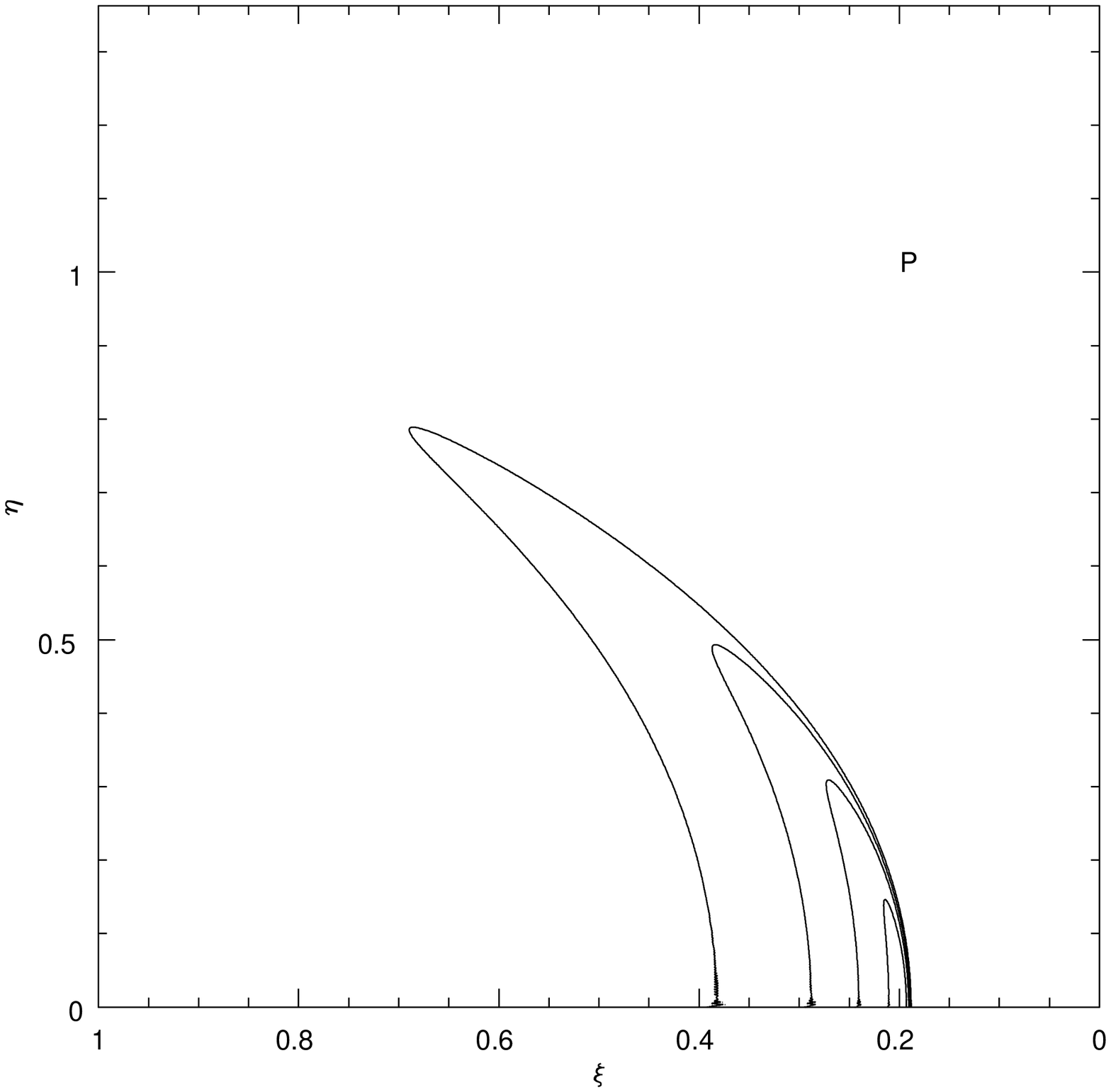}
\caption{{\it Left}: Energy density $E\equiv 4QP$; $E_{\rm max}=2.05$, contours are $E_{\rm max}$ over 2, 4, 8, 16. {\it Right}: Pressure $P$; $P_{\rm max}=0.83$, contours are $P_{\rm max}$ times 0.9, 0.7, 0.5, 0.3. \label{fig1}}
\end{figure}

\begin{figure}
\plottwo{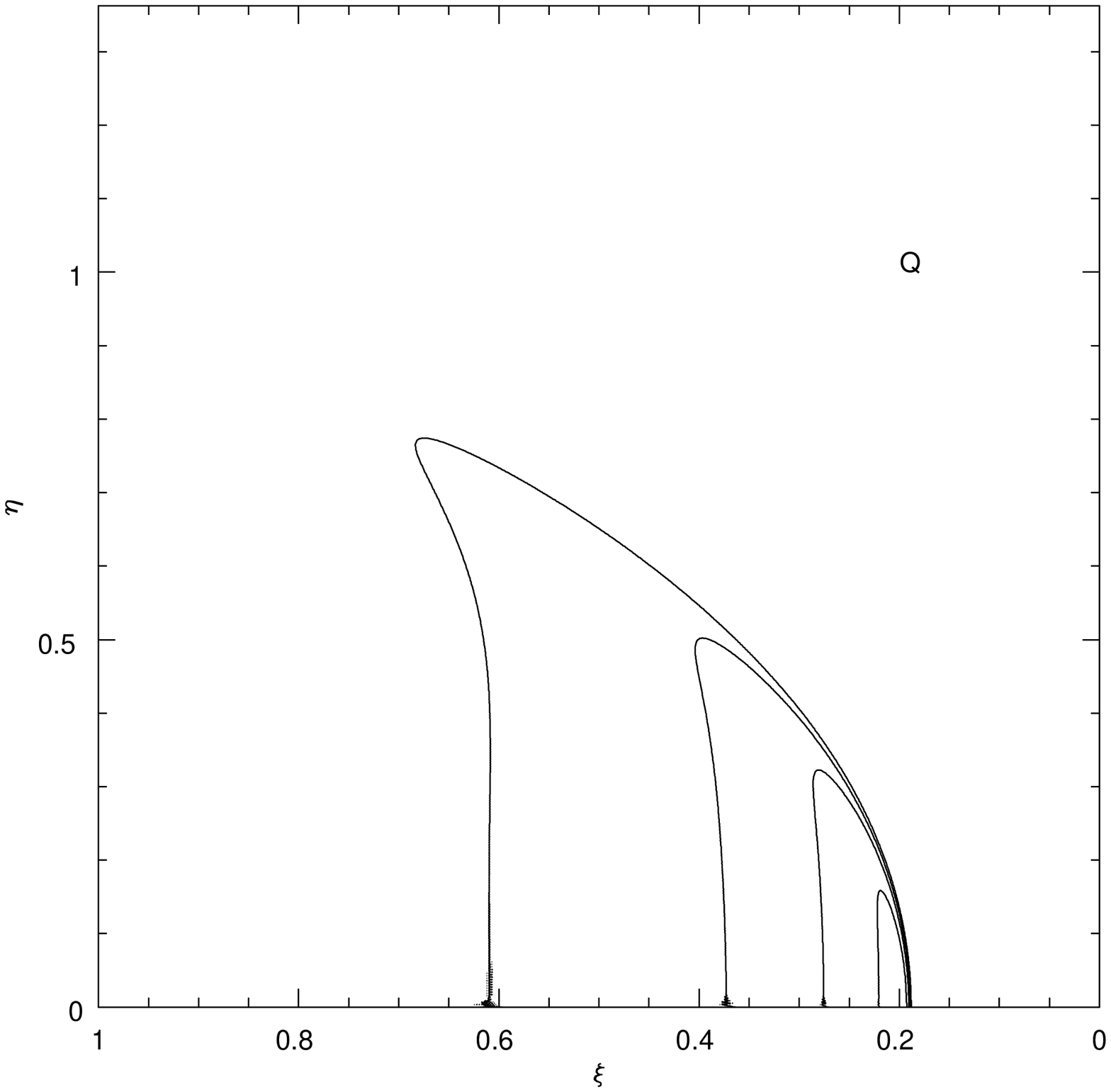}{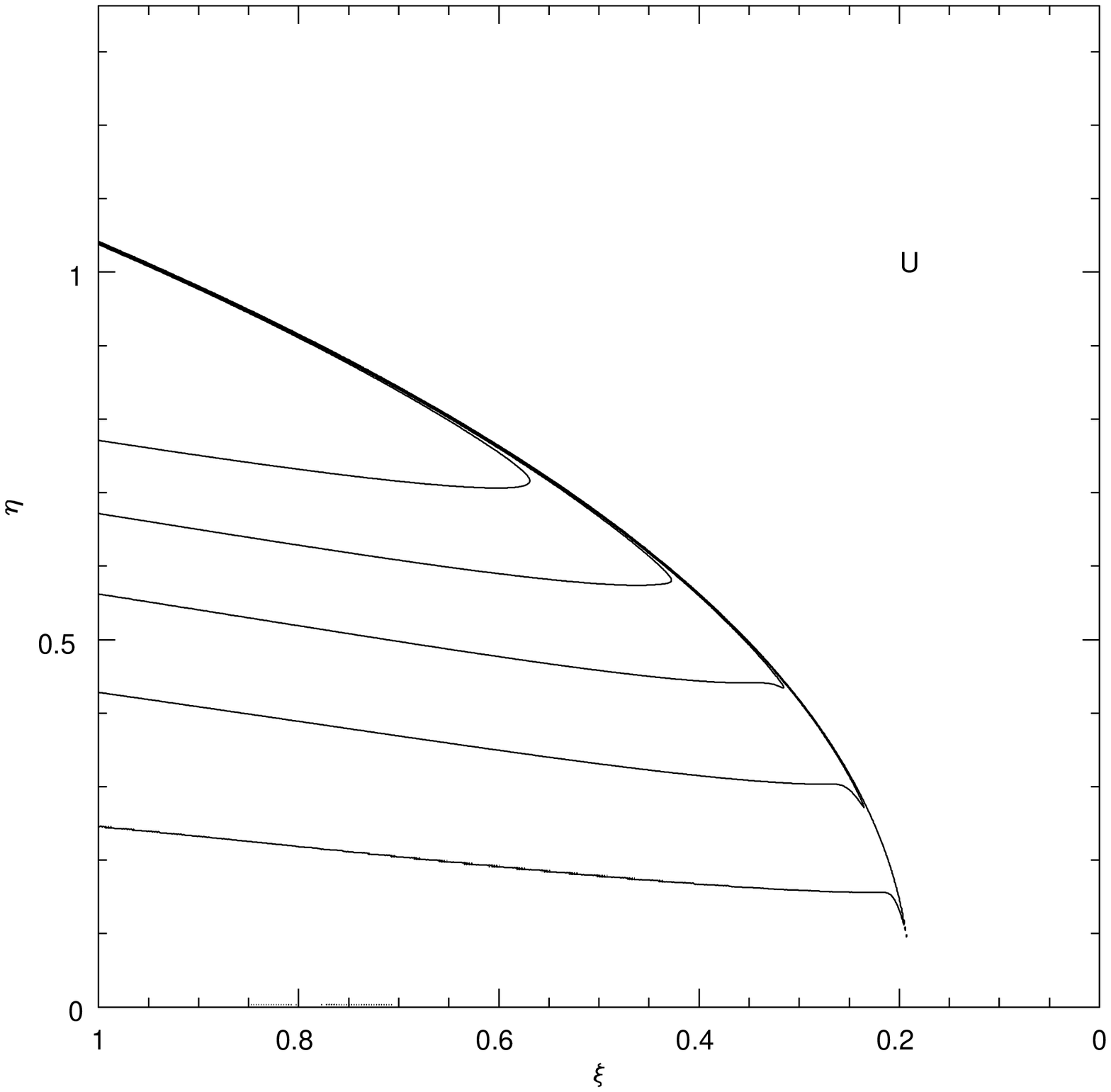}
\caption{{\it Left}: Lorentz factor squared $Q$. ; $Q_{\rm max}=0.63$, contours are $Q_{\rm max}$ times  0.9, 0.7, 0.5, 0.3. {\it Right}: Radial velocity $U$, contours are 1.0, 0.8, 0.6, 0.4, 0.2.\label{fig2}}
\end{figure}

\section{Selfsimilar Directed Ultra-Relativistic Blast Wave}

Selfsimilar directed blast wave is an axisymmetric flow with a shock. The shock is roughly parabolic. The head of the parabola moves at about the speed of light. The energy-containing part of the flow is a narrow shell behind the shock, with a growing opening angle. 

Consider a directed explosion of energy $E$ which sends an ultra-relativistic  blast wave into the medium of uniform density $\rho$. Let $(r,z)$ be cylindrical coordinates, $z$ is the axis of the directed explosion -- meaning that the explosion occurs on the $z$-axis, with the explosion momentum in the positive $z$ direction. 

Then the (proper) pressure $p$, the Lorentz factor squared $q\equiv \gamma ^2$, and the radial (cylindrical radial ) velocity $u$ behind the shock front are given by the following expressions
\begin{equation}\label{at1}
p = e^{-2t/\tau} P(\xi ,\eta )\rho ,~~~ q = e^{-2t/\tau } Q(\xi ,\eta ),~~~ u = e^{t/\tau }U(\xi ,\eta ). 
\end{equation}

Here $P(\xi ,\eta ),~Q(\xi ,\eta ),~U(\xi ,\eta )$ are dimensionless functions of dimensionless variables $\xi ,~ \eta $. Both the dimensionless variables and dimensionless functions are order unity quantities. 

The characteristic time $\tau$ and the dimensionless variables are given by the following expressions 
\begin{equation}\label{at2}
\tau = C\left( {E\over \rho}\right)^{1/3},~~~\xi = e^{-2t/\tau} \left( {t-z\over \tau} \right),~~~ \eta = e^{-t/\tau} \left( {r\over \tau} \right),
\end{equation}
where $C$ is a dimensionless number of order unity. The solution applies for $e^{-t/\tau } \gg 1$, that is for large negative $t$, when equation (\ref{at1}) gives an ultra-relativistic flow (with $q\gg 1$).

We were able to calculate the dimensionless functions only approximately, with some $\sim 10$\% accuracy. The flow fields $P$, $Q$, $U$, are given in figures (1, 2). Due to some inconvenient numerical features of the attractor solution (slow convergence of the energy integral, see \S 3.4), we were able to calculate {\it only the approximate practial value of the dimensionless constant $C$} -- for the initial opening angle of the blast $\sim 0.1$, we find $C\approx 1.5$.

\section{Basic Equations and Computations}

Once the good variables are identified (Blandford and McKee 1976), the computation is straightforward and very similar to the linearized case (Gruzinov 2000). In this section we  

\begin{enumerate}

\item write the relativistic hydrodynamics equations and the shock jump conditions in cylindrical coordinates,

\item simplify the equations assuming the ultra-relativistic flow,

\item find the right self-similar ansatz, and derive the equations describing the selfsimilar functions $P(\xi ,\eta ),~Q(\xi ,\eta ),~U(\xi ,\eta )$,

\item perform the next similarity transformation, and find the asymptotic form of the functions $P(\xi ,\eta ),~Q(\xi ,\eta ),~U(\xi ,\eta )$ in the far downstream region.

\end{enumerate}

\subsection{Relativistic Hydrodynamics in Cylindrical Coordinates}

Relativistic hydrodynamics equations are written as $\partial _\nu T^{\mu \nu}=0$ (Landau \& Lifshitz 1987) . Here $x^\mu=(t,{\bf r})$ are Cartesian coordinates in the rest frame of the unshocked fluid. The energy-momentum is $T^{\mu \nu}=(4u^\mu u^\nu -g^{\mu \nu})p$ in the shocked fluid. The energy-momentum in the unshocked fluid, $\tau ^{\mu \nu}$, has only one non-zero component, and without loss of generality in the final result, we take $\tau ^{00}=1$. The shock position can be represented by the zero isosurface of a scalar field: $\phi=0$ at the shock.   Then the boundary conditions at the shock follow from the hydrodynamics equations:
\begin{equation}\label{shockpar}
T^{\mu \nu }\partial _\nu \phi=\tau ^{\mu \nu }\partial _\nu \phi, ~~~~~~~~~~~\phi=0.
\end{equation}

The 4-velocity is $u^\mu=\gamma (1,{\bf v})$. In cylindrical coordinates, for a cylindrically symmetrical flow, ${\bf v}=v\hat{z} +u\hat{r}$. Use the  three-dimensional form of the hydrodynamics equations 
\begin{equation}
\partial _0T^{00}+\partial _iT^{0i}=0,
\end{equation}
and 
\begin{equation}
\partial _0T^{0i }+\partial _{j}T^{ij}=0,
\end{equation}
and project onto $\hat{z}$ and $\hat{r}$. We get
\begin{equation}\label{cyl1}
\partial _t[(4\gamma ^2-1)p]+4\partial _z[\gamma ^2pv]+4r^{-1}\partial _r[r\gamma ^2pu]=0,
\end{equation}
\begin{equation}\label{cyl2}
4\partial _t[\gamma ^2pv]+\partial _z[(4\gamma ^2v^2+1)p]+4r^{-1}\partial _r[r\gamma ^2pvu]=0,
\end{equation}
\begin{equation}\label{cyl3}
4\partial _t[\gamma ^2pu]+4\partial _z[\gamma ^2pvu]+4r^{-1}\partial _r[r\gamma ^2pu^2]+\partial _rp=0.
\end{equation}

Parameterize the shock front as $\phi\equiv z-z_s(t,r)=0$. Then eq.(\ref{shockpar}) gives:
\begin{equation}
4\gamma ^2p[\partial _tz_s-v+u\partial _rz_s]-p\partial_tz_s=\partial _tz_s
\end{equation}
\begin{equation}
4\gamma ^2v[\partial _tz_s-v+u\partial _rz_s]-1=0.
\end{equation}
\begin{equation}
4\gamma ^2u[\partial _tz_s-v+u\partial _rz_s]+\partial _rz_s=0.
\end{equation}

\subsection{Ultra-Relativistic Hydrodynamics in Cylindrical Coordinates}

Introduce a coordinate grid moving at the speed of light: $x\equiv t-z$ (giving the following replacements: $\partial _t\rightarrow \partial _t+\partial _x$, $\partial _z\rightarrow -\partial _x$). Parameterize the shock position as $z_s=t-x_s(t,r)$. 

Denote $\gamma ^2\equiv q$. Then, 
\begin{equation}
v=1-{1+qu^2\over 2q}-{(1+qu^2)^2\over 8q^2}+O(q^{-3}).
\end{equation}
Here and in what follows we assume $q\gg 1$ and $qu^2=O(1)$. The first condition, $q\gg 1$, means that the flow is ultra-relativistic. The scaling $qu^2=O(1)$ is confirmed by the result.

Keeping the two leading orders in $q^{-1}$, we get
\begin{equation}
\partial _t[(4q-1)p]+\partial _x\left[ \left( 1+2qu^2+{(1+qu^2)^2\over 2q}\right) p\right] +4r^{-1}\partial _r[rqpu]=0,
\end{equation}
\begin{equation}
\partial _t[(4q-2-2qu^2)p]+\partial _x\left[ \left( 1+2qu^2-{(1+qu^2)^2\over 2q}\right) p\right] +r^{-1}\partial _r[r(4q-2-2qu^2)pu]=0,
\end{equation}
\begin{equation}
4\partial _t[qpu]+2\partial _x[(1+qu^2)pu]+4r^{-1}\partial _r[rqpu^2]+\partial _rp=0.
\end{equation}

In the leading order in $q^{-1}$, these can be written in the following final form
\begin{equation}\label{ultra}
\partial _t(qp)+{1\over 4}\partial _x[(1+2qu^2)p]+r^{-1}\partial _r[rqpu]=0,
\end{equation}
\begin{equation}
\partial _t[(1+2qu^2)p]+\partial _x[(1+qu^2)^2q^{-1}p]+2r^{-1}\partial _r[r(1+qu^2)pu]=0,
\end{equation}
\begin{equation}
\partial _tu+{1+qu^2\over 2q}\partial _xu+{u\over 4qp}\partial _xp+{1\over 4qp}\partial _rp=0.
\end{equation}
The boundary conditions, at $x=x_s$:
\begin{equation}\label{ultra1}
q={1\over 4\partial _tx_s+2(\partial _rx_s)^2},~~~p={4\over 3}q, ~~~ u=\partial _rx_s.
\end{equation}

\subsection {Selfsimilar Solution}

The ultra-relativistic blast wave equations (\ref{ultra})-(\ref{ultra1}) admit the following scalings:
\begin{equation}
r\propto t^{1/2}x^{1/2}, ~~~ p\propto q\propto tx^{-1}, ~~~ u\propto t^{-1/2}x^{1/2}
\end{equation}
In $D$ spatial dimensions, the energy of the flow behind the shock corresponding to these scalings is
\begin{equation}\label{ensc}
E\sim xr^{D-1}pq\propto t^{{D+3\over 2}}x^{{D-3\over 2}}.
\end{equation}
During the ultra-relativistic stage the energy of the flow should be close to the explosion energy, meaning that the energy (\ref{ensc}) should be time-independent. This enforces the following scaling for the directed blast wave in $D$ spatial dimensions:
\begin{equation}
x\propto t^{{3+D\over 3-D}}.
\end{equation}
For $D=3$ spatial dimensions this algebraic dependence degenerates into the exponential.

We therefore want to find the following selfsimilar solution of the system (\ref{ultra})-(\ref{ultra1}): 
\begin{equation}\label{ans}
p=e^{-2t}P(\xi,\eta), ~~~ q=e^{-2t}Q(\xi,\eta), ~~~ u=e^tU(\xi,\eta),
\end{equation}
with
\begin{equation}\label{ans1}
\xi=e^{-2t}x, ~~~\eta=e^{-t}r.
\end{equation}
Using the ansatz (\ref{ans}),(\ref{ans1})  in the equations (\ref{ultra})-(\ref{ultra1}), one gets
\begin{equation}\label{ultras}
(-4-2\xi\partial _\xi-\eta \partial _\eta)(QP)+{1\over 4}\partial _\xi[(1+2QU^2)P]+\eta^{-1}\partial _\eta[\eta QPU]=0,
\end{equation}
\begin{equation}
(-2-2\xi\partial _\xi-\eta \partial _\eta)[(1+2QU^2)P]+\partial _\xi[(1+QU^2)^2Q^{-1}P]+2\eta^{-1}\partial _\eta[\eta(1+QU^2)PU]=0,
\end{equation}
\begin{equation}
(1-2\xi\partial _\xi-\eta \partial _\eta)U+{1+QU^2\over 2Q}\partial _\xi U+{U\over 4QP}\partial _\xi P+{1\over 4QP}\partial _\eta P=0.
\end{equation}
The boundary conditions at $\xi=\xi_s(\eta)$:
\begin{equation}\label{ultras1}
Q={1\over 8\xi_s-4\eta \xi_s'+2\xi_s'^2},~~~P={4\over 3}Q, ~~~ U=\xi_s'.
\end{equation}
The explosion energy is
\begin{equation}\label{energy}
E_0=8\pi\int ~d\xi ~\eta d\eta~QP,
\end{equation}
giving the dimensionless constant $C$ of \S2: $C\equiv E_0^{-1/3}$. 

Equations (\ref{ultras})-(\ref{ultras1}) still have a one-parameter scaling group:
\begin{equation}
\xi \rightarrow \lambda ^2\xi ,~~ \eta \rightarrow \lambda \eta ,~~ Q\rightarrow\lambda ^{-2}Q ,~~ P\rightarrow\lambda ^{-2}P, ~~ U\rightarrow\lambda U.
\end{equation}
``Solution'' shown in figures 1, 2 was normalized by the condition $\xi _s(0)=0.2$. This normalization gives order unity values for all other quantities.

The right way to find the attractor is to solve the equations (\ref{ultras})-(\ref{ultras1}). This is not what we have done in this paper. Our ``solution'' was obtained by a direct numerical simulation of the axisymmetric relativistic hydrodynamics equations (\S 4).

\subsection {Selfsimilar Selfsimilar Solution}

Far from the head of the blast wave, at large values of $\eta$, the selfsimilar attractor equations have a selfsimilar solution corresponding to the power law shock position
\begin{equation}\label{sshock}
\xi_s=\eta ^\beta.
\end{equation}

If $\beta \leq 2$, the energy integral (\ref{energy}) diverges. This does not mean that we may automatically dismiss such solution. It might rather indicate that the attractor (\ref{at1}) is never fully filled up. We therefore looked for such solutions. It was found that only the $\beta =2$ solution (with logarithmically divergent energy) exists.

For $\beta \geq 2$, the scaling (\ref{sshock}) enforces the following scalings of the fields:
\begin{equation}
Q=\eta ^{-2(\beta -1)}q(x),~~P=\eta ^{-2(\beta -1)}p(x),~~U=\eta ^{\beta -1}u(x),~~~ x\equiv {\xi\over \eta ^\beta },
\end{equation}
with the boundary conditions on the shock:
\begin{equation}
q(1)={1\over 2\beta ^2}, ~~ p(1)={2\over 3 \beta ^2},~~ u(1)=\beta
\end{equation} 

The functions $q$, $p$, $u$ are obtained from the system of ordinary differential equations:
\begin{equation}\label{ssm}
{1\over 4}[(1+2qu^2)p]'-(3\beta -4)qpu-\beta x(qpu)'=0
\end{equation}
\begin{equation}
[(1+qu^2)^2q^{-1}p]'-(2\beta -4)(1+qu^2)pu-2\beta x[(1+qu^2)pu]'=0
\end{equation}
\begin{equation}\label{ssm1}
2(1+qu^2)u'-2(\beta -1)+(u-\beta x){p'\over p}=0
\end{equation}

Numerical integration of equations (\ref{ssm})-(\ref{ssm1}) then shows that solutions exist only for $\beta <2.032$. {\it But we were unable to determine the true value of $\beta$ theoretically.} The direct numerical simulations (\S 4) give marginal evidence that $\beta >2$ (the case $\beta =2$ has positive $u$ for all $x$, the case $\beta >2$ has negative $u$ for large  enough $x$, we do see negative velocities in the far downstream region). We therefore tentatively conclude that $2<\beta <2.032$. 

Since $\beta$ is so close to the minimal value $\beta =2$, the energy integral (\ref{energy}) converges very slowly. As a result, only extremely directed explosions can reach the pure attractor stage. Explosions with initial opening angles of order 0.1 or 0.01 produce only a partial attractor -- the attractor is being filled up throughout the entire quasi-selfsimilar stage. This should give small (algebraic) corrections to the exponential decay laws of maximal energy and pressure of the pure solution (\ref{at1}). Direct numerical simulations (\S 4) give $\beta \approx 2$, to about 10\% .

\section{Numerical Simulation}

Simple Lax scheme with 800x800 resolution was used to simulate the axisymmetric relativistic hydrodynamics in cylindrical coordinates eq.(\ref{cyl1}-\ref{cyl3}). Coordinate mesh was moving at the speed of light -- this allows to simulate just a small region around the energy containing part of the flow. 

Different initial configurations were tried and seen to produce approximately equal attractors. The solution shown in fig. (1,2) used the following initial condition. We started with a blob of energy moving in the $z$ direction with the maximal Lorentz factor equal to 20. Maximal initial pressure in the blob was 400; the blob was moving into the medium of energy density equal to 1. Maximal initial energy density in the blob was $4pq=640,000$. The simulation was stopped when the maximal energy density of the flow dropped to about $2500$. Initially, the blob had the $r$ to $z$ size ratio equal to 1:20. 

Figures (1,2) show the final state of the blob, in the appropriately rescaled variables. The dimensionless constant $C$ was calculated from the initial energy of the blob and the inferred time constant $\tau$ measured from the numerical solution.

 It was found that the blob does approximately evolve toward the attractor solution given by  (\ref{at1}), (\ref{at2}). Namely (to about 10\% accuracy):  

\begin{enumerate}

\item Maximal energy density $\epsilon _{\rm max}$ of the flow decreased just somewhat faster than exponentially. 

\item Position of the shock $x_s$ grew almost exponentially.

\item $\epsilon _{\rm max}$ and $x_s$ were approximately related by $\epsilon _{\rm max}\propto x_s^{-2}$ in agreement with (\ref{ans}),(\ref{ans1}).

\item Total energy per logarithmic interval of energy density was approximately constant for low energy densities, corresponding to $\beta \approx 2$.
 
\item The time constant $\tau$ inferred from the rate of change of maximal energy density $\epsilon _{\rm max}$, was approximately equal to $\tau =8x_sq_{\rm max}$, in agreement with (\ref{ultras1}).

\end{enumerate}

We should not have expected a better than 10\% agreement. First, Lax method does not give accurate maximal values (which theoretically occur at the shock). At resolution of 400x400, some of the inferred values change by about 10\% . Second, we are simply not ultra-relativistic enough. We stop at the maximal Lorentz factor of about 5. A fair fraction of the flow has too small Lorentz factors. Finally, because of the slow convergence of the energy integral (\S 3), the attractor is being filled up all the time -- hence faster than exponential decrease of the maximal energy density.

\acknowledgements I thank Andrew MacFadyen for discussions. This work was supported by the Davide and Lucile Packard Foundation.

\end{document}